\documentclass[onecolumn]{svjour3}

\usepackage[utf8]{inputenc} 
\usepackage[T1]{fontenc}    
\usepackage{hyperref}       
\usepackage{url}            
\usepackage{booktabs}       
\usepackage{amsfonts}       
\usepackage{nicefrac}       
\usepackage{microtype}      
\usepackage{tikz}
\usepackage{amsmath}
\usepackage{braket}
\usepackage{graphicx}
\usepackage{sidecap}
\usepackage[boxruled,noend,noline]{algorithm2e}
\usepackage{subcaption}
\usepackage{enumitem}
\usetikzlibrary{svg.path, shapes.geometric, arrows}

\tikzstyle{titles} = [rectangle, rounded corners, minimum width=3cm, minimum height=1cm,text centered, draw=black]
\tikzstyle{initialization} = [rectangle, minimum width=3cm, minimum height=1cm, text centered, draw=black, fill=orange!30]
\tikzstyle{measurementCycle} = [rectangle, minimum width=3cm, minimum height=1cm, text centered, draw=black, fill=green!30]
\tikzstyle{arrow} = [thick,->,>=stealth]

\begin{document}

\title{A Decoder for the Color Code with Boundaries}

\author{Skylar Turner \and
        Josey Hanish \and
        Eion Blanchard \and
        Noah Davis \and
        Brian La Cour
}

\institute{ S. Turner \at
            Applied Research Laboratories, The University of Texas at Austin, Austin, TX \\
            \email{skylar\_rt7@utexas.edu}
            \and
            E. Blanchard \at
            University of Illinois at Urbana-Champaign, Champaign, IL
            \email{eionmb2@illinois.edu}
            \and
            J. Hanish \at
            Applied Research Laboratories, The University of Texas at Austin, Austin, TX \\
            \email{josey.hanish@gmail.com}
            \and
            N. Davis \at
            Applied Research Laboratories, The University of Texas at Austin, Austin, TX \\
            \email{noah.davis@arlut.utexas.edu}
            \and
            B. La Cour \at
            Applied Research Laboratories, The University of Texas at Austin, Austin, TX \\
            \email{blacour@arlut.utexas.edu}
}

\date{Received: \today / Accepted: date}

\maketitle


\begin{abstract}
We introduce a decoder for the 3D color code with boundaries, which is a variation of the restriction decoder introduced by Kubicka and Delfosse. Specifically, we adapt the lift procedure to efficiently find a correction on qubits adjacent to a boundary. We numerically estimate a threshold of $4\% - 8\%$ for Pauli $X$ errors and a threshold of $0.7\% - 0.8\%$ for Pauli $Z$ errors. Our work is a first step towards characterizing the performance of Bomb\'{i}n's recently proposed ``colorful quantum computing.''

\end{abstract}


\section{Introduction \label{sec:introduction}}

Quantum computers promise to solve certain problems faster than their classical counterparts \cite{Shor1994,Grover1996}, but the quantum systems used to build one tend to be highly sensitive to errors. Using error-correcting codes it is possible to build a \emph{fault-tolerant} quantum computer, one in which errors are corrected before they can degrade the computation. In such a code, a logical qubit is encoded in many physical qubits, introducing redundancy that makes it possible to correct errors on the physical qubits. In order to correct errors, a \emph{decoding algorithm} must interpret \emph{syndrome} information in order to identify a \emph{correction operator} to remedy the error that has occurred.

The leading candidate for a fault-tolerant universal quantum computing scheme is the surface code with defect braiding and magic state distillation \cite{RaussendorfToric,MartinisToric}. The surface code has a high noise threshold (near 10\% for independent, identically distributed Pauli X and Z errors and about 1\% when measurement errors are included \cite{topologicalQuantumMemory}), below which the likelihood of errors can be made arbitrarily small by using larger numbers of physical qubits. The threshold of the 2D color code has been found to be comparable, also near 10\% \cite{restrictionDecoder}. However, the both of these codes can transversally implement only a limited set of quantum gates and requires magic state distillation \cite{distillationOG} to implement a universal gate set. Magic state distillation, though it has been the subject of an intense research focus, has a very large resource requirement \cite{campbell2017roads}. These drawbacks inspire research into alternative fault-tolerant schemes. ``Colorful quantum computing'' \cite{bombinOG} is one such promising alternative, where a universal gate set is realized by transversal gates, adaptive measurements, and classical processing. Implementing colorful quantum computing requires development of new decoding techniques \cite{bombinOG}

In this paper, we investigate quantum error correction on the 3D color code with boundaries, also called the ``tetrahedral color code,'' which is a necessary structure for colorful quantum computing \cite{bombinOG}. In Section \ref{sec:algorithms}, we introduce an algorithm for decoding Pauli $X$ and $Z$ errors on the tetrahedral color code. This algorithm was adapted from the ``restriction decoder,'' an algorithm for decoding color codes with periodic boundaries \cite{restrictionDecoder}. In Section \ref{sec:results}, we characterize the performance of this decoder on independent, identically distributed $X$ and $Z$ errors and present thresholds for error-correction such that for error rates below the threshold, the probability of logical errors can be made arbitrarily small by using larger size codes. Finally, we interpret our results in relation to colorful quantum computing in Section \ref{sec:conclusion}.

\subsection{Colorful quantum computing}

Colorful quantum computing uses a 3D color code that can be implemented on a 2D lattice of physical qubits using ``just-in-time decoding.'' This scheme achieves universality with only transversal gates, circumventing the Eastin-Knill Theorem \cite{EastinKnillNoGo} by relying on non-local classical computing \cite{bombinOG}. Transversal gates, which operate ``qubit-wise'' on physical qubits, are highly desirable for fault-tolerant codes because they propagate errors only locally \cite{bombinErrorPropagation}. Colorful quantum computing is fault-tolerant, though the noise threshold for fault-tolerance has only been investigated theoretically \cite{bombinOG} and is very very low. However, thresholds found using direct simulation of the error-correction process are usually much higher.

The tetrahedral color code admits the following transversal logical gates \cite{bombinErrorPropagation}: The logical Controlled-$X$ (CNOT) gate can be applied by applying CNOT pairwise between corresponding physical qubits of two logical qubits. The Controlled-$Z$ ($cZ$) gate is applied analogously, but by matching only a subset of qubits, which allows logical qubit cluster states to be arranged in a regular lattice. Most importantly, the $T$ gate is transversal. The logical $T$ gate is applied by applying $T$ to a set of the physical qubits and $T^\dagger$ to the remainder, such that the two sets are a bipartition of the lattice. The logical Pauli $X$ and $Z$ operators are transversal as well, and consist of Pauli $X$ and $Z$, respectively, applied to each physical qubit.

When supplemented with measurements in the $Z$ basis $\{ \ket{0},\ket{1} \}$ and $X$ basis $\{\ket{\pm} = \ket{0}\pm\ket{1}\}$, this set of gates for the tetrahedral color code becomes universal in measurement-based quantum computing (MBQC) \cite{oneWayQC,ftOneWayQC}.  TO implement a non-Pauli operation, an $X+Y$ measurement $\{ \ket{0}\pm e^{i\pi/4}\ket{1} \}$ can be performed by applying a T gate then measuring in the $X$ basis. An example of how to implement the Hadamard gate in such a scheme can be found in Ref.\ \cite{jozsaIntroMBQC}.

MBQC is equivalent to the circuit model, but implemented differently - all entanglement is present in a resource state at the beginning of the computation. Gates are simulated by qubit measurements on a highly entangled \emph{cluster state}, and a classical computer processes the Pauli frame \cite{oneWayQC}. Universality requires a cluster state of at least two dimensions and the ability to measure qubits in the $X$, $Z$, and $X+Y$ bases \cite{jozsaIntroMBQC}. Generating a cluster state requires a source of $\ket{+}$ states and the ability to apply the $cZ$ gate between neighboring qubits \cite{jozsaIntroMBQC}.

Colorful quantum computing is a type of MBQC that encodes the logical qubits of the cluster state in a tetrahedral color code. Arranged in this way, the initial state forms a 3D lattice of qubits. The cluster state is initialized using $cZ$ gates, $\ket{+}$ states, and ancilla qubits. Afterwards, the logical qubits form a cluster state up to known single-qubit Pauli errors, i.e. the Pauli frame \cite{ftOneWayQC}. An $X$-error decoder is used to determine the Pauli frame and the locations of errors are stored in a classical memory. The measurement pattern is then enacted by measuring the logical qubits in the appropriate basis, which requires either an $X$-error decoder, a $Z$-error decoder, or both. If the logical measurement is in the $X\!+\!Y$ basis, a transversal $T$ gate is applied before measuring in the $X$ basis, which requires a $Z$-error decoder to interpret. This scheme does not require magic state distillation because the $cZ$ and $T$ gates are transversal, as are measurements in the Pauli bases. The ability of colorful quantum computing to realize fault-tolerant, universal quantum computing without magic state distillation motivates our development of decoders for the tetrahedral color code.

\subsection{Error correction in the stabilizer formalism}

3D color codes are a type of error-correcting stabilizer code. In the stabilizer formalism, states are described not by a wavefunction but by stabilizer operators. The state described by a set of stabilizer operators $\mathcal{S}$ is the $+1$ eigenstate of each operator, $S\!\ket{\psi}=\ket{\psi}, S\in \mathcal{S}$. A state of $n$ physical qubits described by $k$ independent, commuting stabilizers inhabits a $2^{n-k}$ dimensional subspace. In this way, we say there are $n-k$ \emph{logical} qubits. Operators that commute with the set of stabilizers are denoted $N(\mathcal{S})$, and the set of non-trivial logical operators is $N(\mathcal{S}) \setminus \mathcal{S}$. In the context of stabilizer codes, the code space is the $2^{n-k}$ dimensional $+1$ eigenstate of the code stabilizers. To define a basis for the $n-k$ logical qubits, one chooses $n-k$ logical $\bar{Z}_i$ operators from $N(\mathcal{S}) \setminus \mathcal{S}$. Then one can choose logical $\bar{X}_i$ operators such that $\bar{X}_i\bar{Z}_j = (-1)^{\delta_{ij}}\bar{Z}_j\bar{X}_i$. 


An error, $E$, will anticommute with at least one stabilizer if the error is not a logical operator. This results in a $-1$ measurement outcome when one of those stabilizers is measured. Thus, we measure a generating set of stabilizers to find the \emph{syndrome} corresponding to the error $E$, which we then decode to obtain the corresponding \emph{correction operator}, $\tau$. Finally, we apply this correction operator to bring the system back into the code space. Error correction fails when the combination of errors and correction operators is a non-trivial logical operator; i.e., $\tau E \not\in N(\mathcal{S}) \setminus \mathcal{S}$.

\subsection{Structure of the color code}

3D color codes are topological stabilizer codes. These codes are defined on lattices in which stabilizers correspond to topological objects, such as loops or surfaces. An accessible introduction to color codes (of all dimensions) may be found in Ref.\ \cite{niceIntroColorCodes}. Here, we investigate \emph{tetrahedral} color codes, which are 3D color codes with non-periodic boundary conditions \cite{BombinMartinDelgado2007}.

\begin{figure}[h]
\includegraphics[width=0.48\linewidth]{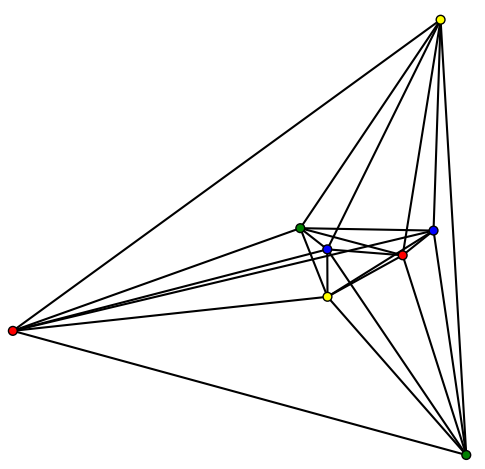}\quad\includegraphics[width=0.48\linewidth]{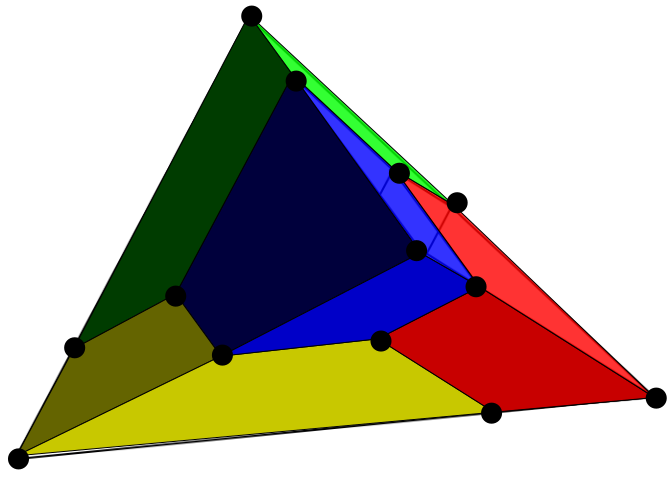}
\caption{(Color online) A representation of the smallest tetrahedral color code, in the dual (left) and primal (right) lattice. Colored vertices in the dual correspond to colored cells in the primal, except for the four \emph{quasivertices} furthest from the center}\label{fig:k1_lattices}
\end{figure}

We define the tetrahedral color code on a dual lattice $\mathcal{L}^\ast$ such that tetrahedra (3-simplices, denoted $\Delta_3(\mathcal{L}^\ast)$) specify the physical qubits, vertices (0-simplices, $\Delta_0(\mathcal{L}^\ast)$) generate the set of $X$ stabilizers, $S_X$, and edges (1-simplices, denoted $\Delta_1(\mathcal{L}^\ast)$) generate the set of $Z$ stabilizers, $S_Z$. The tetrahedral color code depicted in Fig.  \ref{fig:k1_lattices} has 15 qubits, four $X$ stabilizers, and 18 $Z$ stabilizers. 

Each vertex of $\mathcal{L}^\ast$ is assigned a color from the set $\mathcal{C} = \{r,\ g,\ b,\ y\}$ such that no adjacent vertices share a color. Each edge is labeled by the two colors that are absent from the vertices it connects. For example, a $by$-colored edge connects an $r$-colored vertex and a $g$-colored vertex. 

We refer to the four boundary vertices of the code as quasivertices. The quasivertices do not correspond to $X$ stabilizers, and the tetrahedron formed by connecting all the quasivertices does not correspond to a physical qubit. However, the edges and tetrahedra adjacent to the quasivertices are $Z$ stabilizers and physical qubits, respectively.

For the purposes of physical implementation, it is simpler to define the tetrahedral color code differently. Before, we used the dual lattice description, $\mathcal{L}^\ast$, and now we introduce the primal lattice description, $\mathcal{L}$. In the primal lattice, qubits are vertices, $Z$ stabilizers are faces, and $X$ stabilizers are cells. The four boundaries now appear as triangular facets.

In the simplest possible primal lattice, shown in Fig. \ref{fig:k1_lattices}, each face is adjacent to four qubits and each cell is adjacent to eight qubits. It is harder to see, but in the corresponding dual lattice each edge and vertex is adjacent to four or eight tetrahedra, respectively. We analyze a family of codes built in the dual on the body-centered cubic, or \emph{bcc}, lattice. In the corresponding primal lattice each cell in the bulk is a bitruncated cube. See \cite{BombinNJP2015} for pictures of the lattice we use and a method for generating it.

Every tetrahedral color code encodes a single logical qubit in an entangled state of a larger number of physical qubits. A common basis for a single logical qubit is one in which the logical operators $\bar{X}$ and $\bar{Z}$ are tensor products of $X$ and $Z$ operators, respectively, on each of the physical qubits.


\section{Decoding algorithms \label{sec:algorithms}}

\subsection{$X$ errors --- loop-like syndromes}

We implement the cellular automaton-based restriction decoder described in \cite{restrictionDecoder}, adapting it to a lattice with a tetrahedral boundary (as opposed to a lattice with periodic boundaries). The syndrome of an $X$ error appears as a collection of edges that bound the set of tetrahedra corresponding to the qubits that have errors. Given a set of edges $\sigma \subset \Delta_1(\mathcal{L}^\ast)$, we find the set of tetrahedra $\tau \subset \Delta_3(\mathcal{L}^\ast)$ such that $\sigma = \partial _{3, 1}(\tau)$, where $\partial _{n, m}$ denotes the projection of an $n$-dimensional set of objects to its $m$-dimensional boundary.

The restriction decoder achieves this via two routines: the ``sweep’’ decoder and the ``lift’’ routine. One of the four lattice colors is chosen as the lift color; the rest are sweep colors. For each sweep color, the decoder considers the subset of edges in $\sigma$ labeled with the particular sweep color (i.e. only syndrome edges not incident to a sweep-colored vertex are considered). The decoder then runs the sweep decoder \cite{sweepDecoder}, a toric-code decoder, on this subset of edges to obtain a set of faces whose boundary is the set of edges having been considered. The faces resulting from each sweep color are taken in union as $\gamma \subset \Delta_2(\mathcal{L}^\ast)$. Then $\partial_{2,1}(\gamma) = \sigma$.

Next, considering the subset of vertices assigned the lift color, the decoder identifies the set of tetrahedra $\tau \subset \Delta_3(\mathcal{L}^\ast)$ such that $\partial_{3,2}(\tau) \supset \gamma$ and for each lift vertex $v$, $\partial_{3,2}(\tau)\big|_{v} = \gamma\big|_v$, where $\gamma\big|_v$ is the subset of $\gamma$ containing $v$.  We note the na\"ive treatment of boundary vertices in the lift process. Lifting each bulk (interior) vertex runs in constant time since there are $2^{12}$ possible subsets of tetrahedra incident to each bulk vertex (a consequence of the lattice geometry).  With non-periodic boundaries as in our code lattice, however, the neighborhoods of boundary vertices become increasingly dense at a quadratic asymptotic rate $O(d^2)$, where $d$ is the code distance. So our na\"ive approach runs in at-worst exponential time: $O\big(2^{d^2}\big)$ 

\begin{figure}[t]
\includegraphics[width=0.9\linewidth]{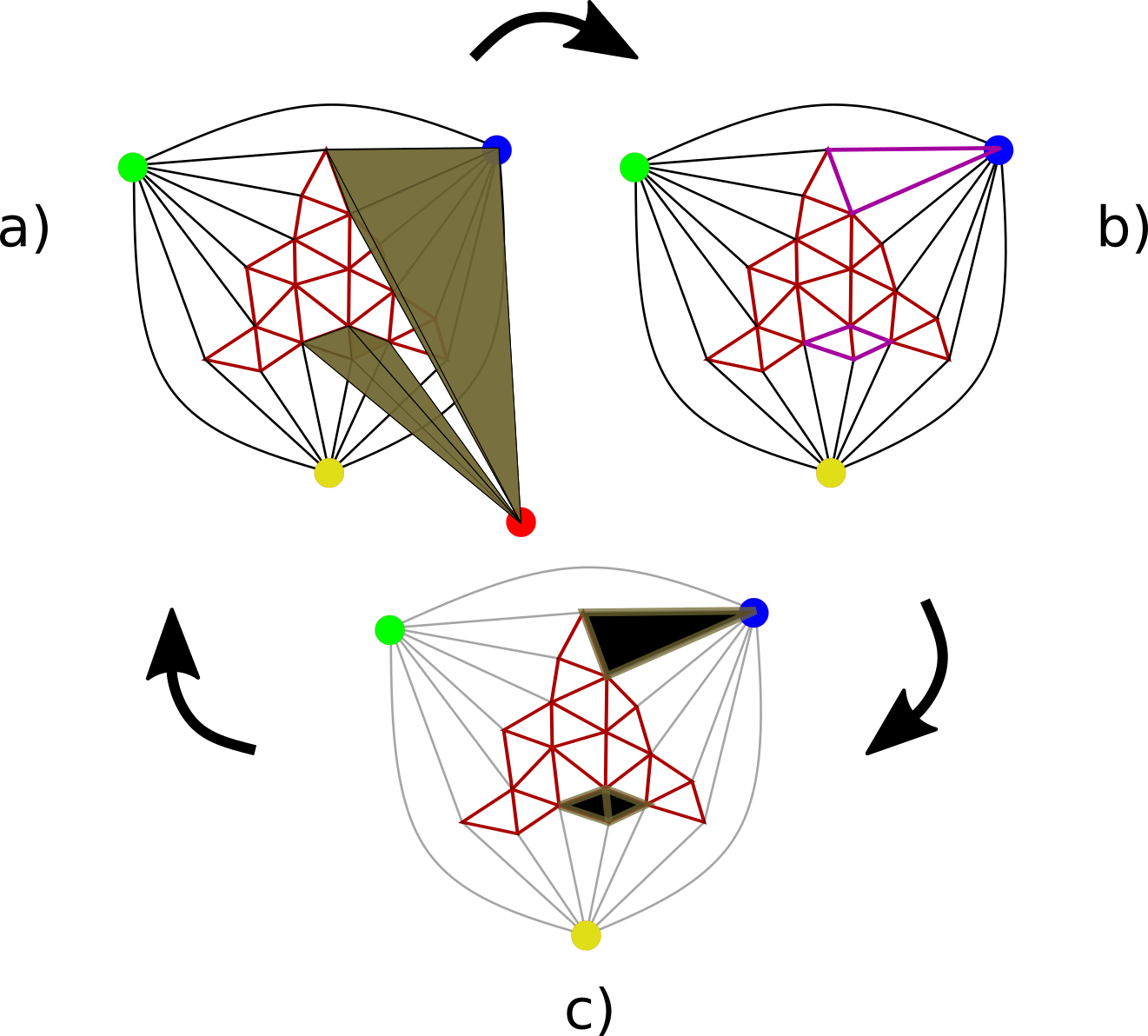}
\caption{(Color online) Illustration of the efficient lift procedure on a boundary quasivertex. In a), lift must find which tetrahedra in the neighborhood of the red quasivertex have a boundary of faces that match the brown highlighted ones. b) We project the marked faces onto the facet nearest the red quasivertex, marking a purple edge for each face's intersection with the facet. c) On the facet, it is possible to quickly find the 2D boundary of these edges marked in black. These faces have a one-to-one correspondence with the error qubits adjacent to the red quasivertex.}
\label{fig:boundaryLift}
\end{figure}

To overcome this computational hurdle, we adapt the ``peel’’ algorithm described in \cite{ourSaviors} to lift boundary vertices more efficiently (See Fig. \ref{fig:boundaryLift} and Alg. \ref{alg:peel}).  This method is an adaptation of the peel decoder originally introduced for the 2d surface code \cite{Delfosse2020}. In fact, on code lattices with small distance, this is more efficient than the na\"ive lift on even the bulk vertices. The peel algorithm proceeds on each lift vertex $v$ by taking the set of tetrahedra $\tau|_v$ incident to $v$ and considering $\partial_{3,2}(\tau|_v) \setminus \big(\partial_{3,2}(\tau|_v)\big|_v\big)$. That is, we consider the set of faces which are incident to $\tau|_v$ but disjoint from $v$ itself (i.e. faces whose vertices are in the \emph{link} of $v$); these faces form a topological sphere when $v$ is a bulk vertex and form a triangular facet when $v$ is a boundary vertex.

We consider the subset of ``intermediate syndrome'' faces which is output from the sweep decoder) that are incident to $v$: $\gamma\big|_v$ (brown faces in Fig. \ref{fig:boundaryLift}a). We project this set of faces $\gamma\big|_v$ into a set of edges (purple in Fig. \ref{fig:boundaryLift}b) on the aforementioned ``peeling surface'' (topological sphere or tetrahedral facet). It is necessary for the peeling surface to be $2$-dimensional, so a face is removed when the algorithm runs on a bulk vertex. The peel algorithm finishes by identifying the set of faces (Fig. \ref{fig:boundaryLift}c) on the surface whose edge-boundary matches these projected edges, which can be done efficiently following Ref. \cite{ourSaviors}, then projects the triangle faces back up to tetrahedra, denoted $\tau\big|_v$, with the associated lift vertex $v$ as the fourth vertex for each tetrahedra. The resulting set of tetrahedra from all lift vertices is the desired $\tau$, i.e. $\tau=\bigcup_{v}\tau\big|_v$

\begin{algorithm}
\caption{Lift subroutine on single vertex}\label{alg:peel}
\SetKwInOut{Input}{input}\SetKwInOut{Output}{output}
\SetKwData{BoundaryFaces}{F}\SetKwData{SyndromeEdges}{SE}\SetKwData{BoundaryEdges}{E}\SetKwData{Tetrahedra}{T}\SetKwData{CurrentEdge}{current\_edge}\SetKwData{CurrentFace}{current\_face}\SetKwData{FaceEdge}{fe}\SetKwData{CorrectionFace}{cf}\SetKwData{CorrectionFaces}{correction\_faces}\SetKwData{CorrectionTetrahedra}{$\tau\big|_v$}
\Input{lift vertex ,$v$; intermediate syndrome faces adjacent to $v$, $\gamma\big|_v$; dual lattice,$\mathcal{L}^\ast$}
\Output{correction tetrahedra, $\tau\big|_v$}
\BlankLine/* Initialization */

\Tetrahedra$\leftarrow$ tetrahedra in $\mathcal{L}^\ast$ containing $v$\;
\BoundaryFaces $\leftarrow$ faces of \Tetrahedra not containing $v$\;
\BoundaryEdges $\leftarrow$ edges in \BoundaryFaces\;
\SyndromeEdges $\leftarrow$ edges in $\gamma\big|_v$ that are also in \BoundaryFaces \;
\If{v is in the bulk}{
  remove any face from \BoundaryFaces}
\BlankLine/* Peel loop, following Ref. \cite{ourSaviors} */

\While{\BoundaryFaces is not empty}{
  \CurrentEdge $\leftarrow$ any edge in \BoundaryEdges that belongs to only one face in \BoundaryFaces\;
  \CurrentFace $\leftarrow$ face that \CurrentEdge belongs to\;
  \If{\CurrentEdge in \SyndromeEdges}{
    add \CurrentFace to \CorrectionFaces\;
    \For{\FaceEdge $\in$ all edges of \CurrentFace in \SyndromeEdges}{
      \eIf{\FaceEdge is already in \SyndromeEdges}{
        remove \FaceEdge from \SyndromeEdges}{
        add \FaceEdge to \SyndromeEdges}
    }
    remove \CurrentEdge from \BoundaryEdges\;
    remove \CurrentFace from \BoundaryFaces\;
  }
}
\BlankLine
\If{\SyndromeEdges is not empty}{
  \Return Failure}
\BlankLine/* Projection of \CorrectionFaces back to tetrahedra */

\For{\CorrectionFace in \CorrectionFaces}{
  add (\CorrectionFace$+$\{v\}) to \CorrectionTetrahedra}
\If{there are more tetrahedra in \CorrectionTetrahedra than in $T-$\CorrectionTetrahedra}{
  \CorrectionTetrahedra $=T-$\CorrectionTetrahedra }
\BlankLine
\Return \CorrectionTetrahedra
\end{algorithm}

The peel subroutine runs in asymptotically constant time for bulk vertices. Its runtime on boundary vertices is now $\mathcal{O}(d^2)$, a vast improvement over the original lift procedure. In addition, decoding is done locally.  

\subsection{$Z$ errors --- point-like syndromes}

The syndrome of $Z$ errors is a set of vertices corresponding to $X$ stabilizers that returned a $-1$ measurement outcome. We use a minimum-weight perfect matching (MWPM) subroutine to find edges that connect these vertices \cite{topologicalQuantumMemory,edmonds1969blossom}. First, we restrict the dual lattice to include only vertices of two colors; that is, for $\mathcal{C}$ as above, vertices assigned color $\kappa_i$ or $\kappa_j$ are removed, while vertices assigned $\kappa_k$ or $\kappa_l$ remain (for $\kappa_i \neq \kappa_j \neq \kappa_k \neq \kappa_l$). We then match the syndrome vertices on each twice-restricted lattice and take the union of all returned sets of edges. Those edges correspond to the syndrome that would occur if $X$ errors had occurred on the qubits actually affected by $Z$ errors. Finally the $X$-error decoder, described above, is called as a subroutine to identify the error qubits. This is the approach used in \cite{aloshiusSarvepalliColorCode}, which can be thought of as a specific implementation of the more general process described by the restriction decoder \cite{restrictionDecoder}.

To handle the boundary, we adapt the MWPM algorithm to allow for vertices to match to the nearest quasivertex that remains in the twice-restricted lattice. We add edges between each vertex and its nearest quasivertex, which can be either color. In addition, we add weight-$0$ edges between quasivertices so that the solutions that do not include a quasivertex-vertex edge are considered to be matchings. This is the same process that is used to decode the surface code using MWPM \cite{boundariesMWPM}, and it is not computationally any more difficult than decoding without boundaries. Our decoding scheme corrects the smallest set of error qubits that could have caused the given syndrome. This is a justified choice since the probability $p^n$ of an $n$-qubit error decreases as $n$ increases; smaller errors are more likely. Because of this, the decoder has a tendency to match bulk vertices to the boundary quasivertices, while an observer who knows the positions of the original error qubits would expect that vertex to be matched to a bulk vertex. This does not change the effective distance of our code; the smallest error that the decoder turns into a logical error still has weight $(d-1)/2$. It may be possible to weight edges that connect quasivertices to bulk vertices to enhance performance, similar to the flag qubit scheme used in Ref.\ \cite{kubickaScoop}. We leave more sophisticated ways to handle Z boundary errors for future research.

The $Z$-error decoder is necessarily less successful than the $X$-error decoder: there are fewer $X$ stabilizers than there are $Z$ stabilizers, and $X$ stabilizers have higher weight than $Z$ stabilizers, so $Z$-error syndromes contain less information than $X$-error syndromes. The $Z$-error decoder also has a worse runtime than the $X$-error decoder because of the repeated MWPM subroutines.

\section{Characterization of performance \label{sec:results}}

We numerically tested both the $X$-error decoder and the $Z$-error decoder under an assumption of independent and identically distributed local noise. Each data point was found using between $10^4$ and $10^7$ Monte Carlo simulations. Error bars represent the standard deviation of the mean, given by $\sqrt{p_{\text{fail}} (1-p_{\text{fail}})/N}$, where $p_{\text{fail}}$ is the decoding failure probability and $N$ is the number of Monte Carlos simulations used. On many of the data points it is not possible to see the error bars, as they are very small.

\begin{figure}[t]
    \centering
    \begin{subfigure}[t]{0.8\textwidth}
        \includegraphics[width=0.98\linewidth]{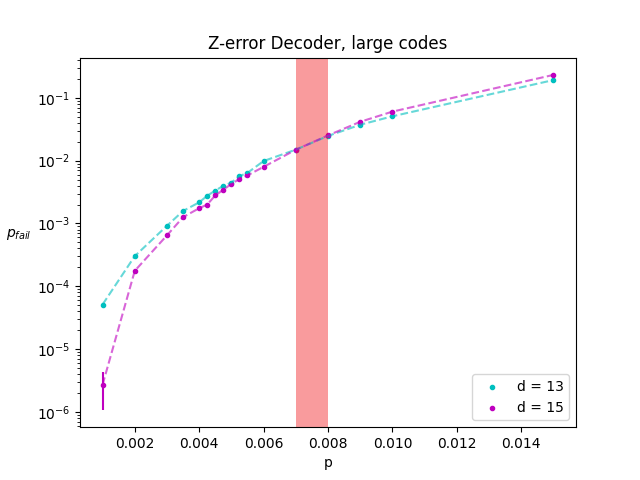}
        \caption{(Color online) $Z$-error threshold is between $0.7\%$ and $0.8\%$}
        \label{fig:z_big}
    \end{subfigure}
    \\
    \begin{subfigure}[t]{0.8\textwidth}
    \includegraphics[width=0.98\linewidth]{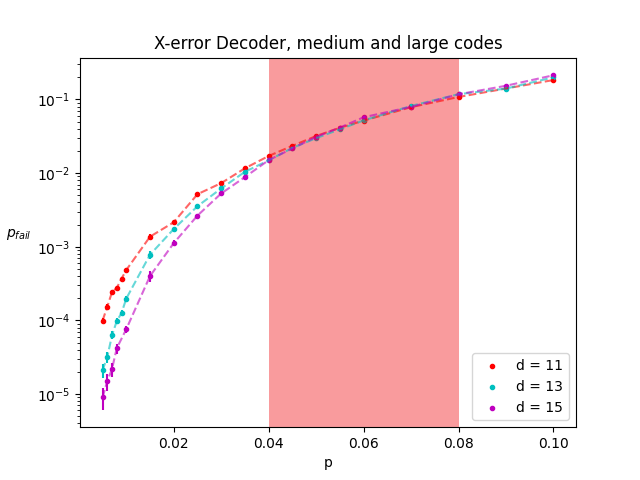}
    \caption{(Color online) $X$-error threshold is between $4\%$ and $8\%$ }
    \end{subfigure}
    \caption{Thresholds}\label{fig:thresholds}
\end{figure}

\begin{figure}[t]
    \centering
    \includegraphics[width=0.98\linewidth]{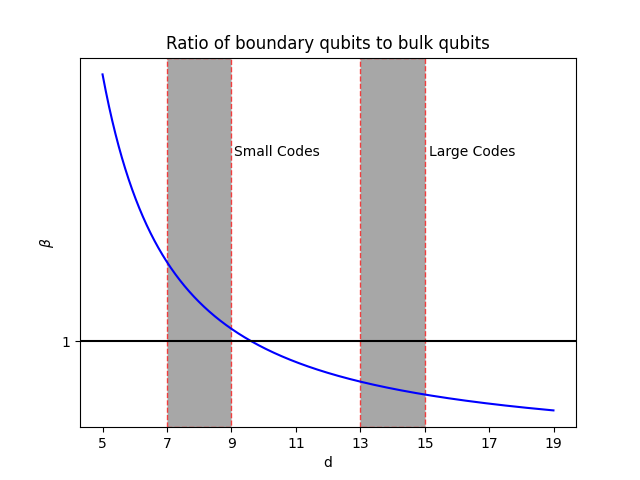}
    \caption{(Color online) Ratio of boundary qubits to bulk qubits,$\beta$, in tetrahedral color codes}
    \label{fig:bb_ratio}
\end{figure}

\begin{figure}[t]
    \begin{subfigure}[t]{0.8\textwidth}
        \includegraphics[width=0.98\linewidth]{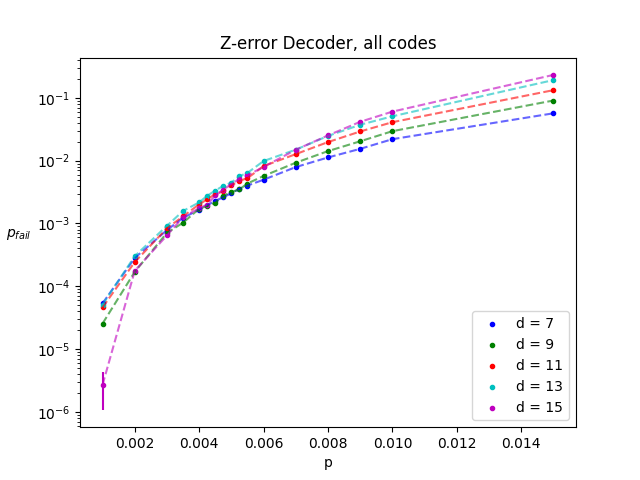}
        \caption{(Color online) $Z$-error data}
        \label{fig:z_all}
    \end{subfigure}
    \\
    \begin{subfigure}[t]{0.8\textwidth}
        \includegraphics[width=0.98\linewidth]{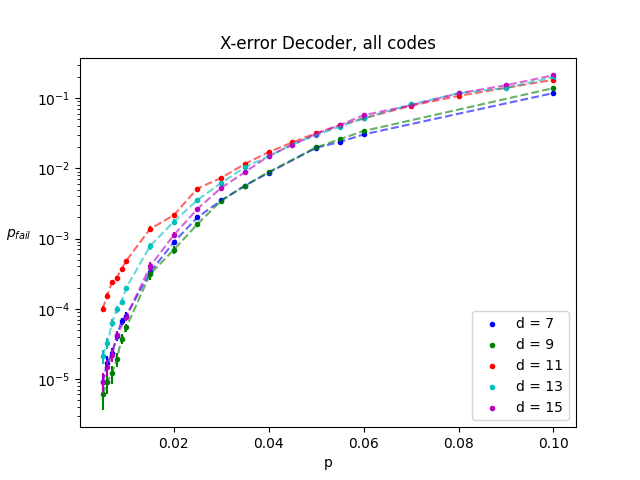}
        \caption{(Color online) $X$-error data}
        \label{fig:x_all}
    \end{subfigure}
    \caption{All data}\label{fig:all}
\end{figure}

 We found that the performance of the decoder seemed to correlate with the ratio of boundary qubits to bulk qubits, denoted $\beta$, in each code block. For example, smaller codes, where there are more boundary qubits than there are bulk qubits (i.e., $\beta>1$) crossover at a different error probability than the larger codes, where bulk qubits outnumbered the boundary qubits. In other words, the data points to different \emph{pseudo-thresholds}. We must decide which crossover point is the true threshold. It is a common assumption that the threshold for an error-correcting code with boundaries should be the same as the similar error-correcting code without boundaries. This is true, for example, with the thresholds of the surface code and the toric code \cite{boundariesMWPM}. This is because as the codes grow larger, the number of qubits on the boundary becomes small compared to the number of qubits in the bulk. This reasoning still holds in the case of 3-d color codes, as $\beta=O(\frac{1}{x})$. Interestingly, figures 10 and 11 from Ref. \cite{boundariesMWPM} show this dependence on $\beta$ in numerical tests of the surface code's threshold. The authors attribute the multiple pseudo-thresholds to boundary effects, at least one of which is caused by the surface code boundary stabilizers being lower-weight than the bulk stabilizers. Analysis using the $\beta$ idea agrees with their data, as the only code which intersects in a different place from the other ones is the $d=3$ surface code, where $\beta>1$. Codes for which $d\geq 5$ have $\beta<1$ for the surface code, and they cross the performance threshold together. It can be seen in their data that the pseudo-thresholds converge to the toric code's threshold as $\beta\rightarrow 0$.
 
 The difficulty in simulating error correction on large tetrahedral color codes makes it hard to test on large enough lattices whether the boundary is unimportant. (In Ref. \cite{gaugeColorCodeTest} the authors simulated error correction on the related 3-d gauge color code and were able to test large code sizes with a less computationally complex decoder.) For the codes plotted in Fig. \ref{fig:z_big}, $\beta<1$, which points to the actual threshold for Z-error correction being between $p=0.7\%$ and $p=0.8\%$. This agrees with the results from Ref. \cite{restrictionDecoder}, which found a threshold of $0.77\%$ for $Z$ errors for the 3d color code without boundaries. For the $X$-error decoder, the threshold is between $4\%$ and $8\%$ which is a little less than the 13.1\% threshold estimated in Ref. \cite{restrictionDecoder} for the 3d color code without boundaries. 
 
 The difference in performance between differently sized codes can be seen when all the data is plotted together. There is a clear separation between the $d=7,9$ codes, where $\beta>1$, and the $d=11,13,15$ codes, where $\beta<1$, shown in Fig. \ref{fig:x_all}, and a similar separation exists for the $Z$-error data (Fig. \ref{fig:z_all}) though it is less pronounced. The $\beta>1$ codes having a lower pseudothreshold than the $\beta<1$ codes makes clear the fact that the boundary plays a very important role in error-correction of the tetrahedral color code.

\section{Conclusion \label{sec:conclusion}}

We are able to adapt the restriction decoder to boundaries, without much reduction in performance. The highest possible values for correction of $X$ errors and $Z$ errors in a 3D color code with periodic boundaries is $27.6\%$ and $1.9\%$ respectively \cite{statMechThresh}.  Our decoder achieves thresholds an order of magnitude below these. Previous tests of the restriction decoder on the \emph{bcc} lattice without boundaries found a threshold of $0.77\%$ for the $Z$-error decoder, and estimated (though not numerically tested) a threshold of $13.1\%$ for the $X$-error decoder \cite{restrictionDecoder}. The $Z$-error decoder threshold of $0.46\%$ found in \cite{gaugeColorCodeTest} using a ``clustering decoder,'' to our knowledge, is the only previous threshold found for a 3D color code with boundaries (though it was not tested on the \emph{bcc} lattice).

We found that an experimentally-determined threshold depends on the size of the codes used in numerical simulation. The threshold found from testing our large codes agrees with the threshold found by testing similar codes without boundaries, which seems to suggest that as long as $\beta>0$ the threshold will be the same as $\beta=0$. It would be an interesting avenue of research to understand this relationship more fully.

The success of adapting the restriction decoder into a fault-tolerant protocol for the 2D color code in Ref. \cite{kubickaScoop} inspires hope that similar approaches may be used to improve performance of the tetrahedral color code in a fault-tolerant setting. Tetrahedral color codes are the building blocks of colorful quantum computing, which can be implemented using only transversal gates and a 2D lattice of physical qubits. Our work is a first step towards being able to numerically assess the performance of colorful quantum computing.


\bibliographystyle{spphys}
\bibliography{references}


\end{document}